\documentstyle[eqsecnum,aps,epsf]{revtex} 

\begin{document}

\twocolumn[\hsize\textwidth\columnwidth\hsize\csname@twocolumnfalse\endcsname

\title{Coupled Bose-Einstein condensate: Collapse  for
attractive interaction}

\author{Sadhan K. Adhikari}
\address{Instituto de F\'{\i}sica Te\'orica, Universidade Estadual
Paulista, 01.405-900 S\~ao Paulo, S\~ao Paulo, Brazil\\}

\date{\today}
\maketitle
\begin{abstract}

We study the  collapse in a coupled Bose-Einstein condensate
of two types of bosons 1 and 2 under the action of a trap using the
time-dependent Gross-Pitaevskii equation. The system may undergo collapse
when one, two or three of the scattering lengths $a_{ij}$ for scattering
of boson $i$ with $j$, $i,j = 1, 2 $, are negative representing an
attractive interaction. Depending on the parameters of the problem a
single or both components of the condensate may experience collapse.

{\bf PACS Number(s): 03.75.Fi,05.30.Jp}

\end{abstract}

\vskip1.5pc]

\section{Introduction}
The experimental detection \cite{1,1a} of Bose-Einstein condensation (BEC) 
at ultralow temperature in dilute trapped bosons (alkali metal and
hydrogen atoms and the recent possibility in  molecules)  has spurred
intense
theoretical activities on various aspects of the condensate
\cite{2,3,4,5,6}.  Many properties of the condensate are usually
described by the mean-field time-dependent Gross-Pitaevskii (GP)
equation
\cite{5,8}. One of the most interesting features of BEC has been observed
in the case of attractive interatomic interaction \cite{1a,2}. In that
case
the condensate is stable for a maximum critical number of atoms, beyond
which the condensate experiences a collapse. When the number of atoms
increases beyond the critical number, due to interatomic attraction the
radius tends to zero and the central density of the condensate tends to
infinity. Consequently, the condensate collapses emitting particles until
the number of
atoms is reduced below the critical number and a stable configuration is
reached. The condensate may experience a series of collapses \cite{1a,2}.
This phenomenon was observed in the BEC of $^7$Li atoms with negative
scattering length denoting attractive interaction where the critical
number of atoms was about 1400 \cite{1a,2}. Theoretical analyses based on
the GP equation in the case of $^7$Li atoms also confirmed this collapse
\cite{1a,2,5,6}.

More recently, there has been experimental realization of BEC involving
atoms in two different quantum states \cite{excpl1,excpl2}. In one
experiment
$^{87}$Rb atoms formed in the $F=1$, $m=-1$ and $F=2$, $m=1$ states by the
use of a laser served as two different species, where $F$ and $m$
are the total angular momentum and its projection \cite{excpl1}. In
another experiment a coupled BEC was formed with the $^{87}$Rb atoms in
the $F=1, m=-1$ and $F=2, m=2$ states
\cite{5,excpl2}.  It is possible to use the same magnetic trap to confine
atoms in two magnetic states and this makes these experimental
investigations technically simpler compared to a realization of BEC with
two different types of atoms requiring two different trapping mechanisms.
This is why so far it has not been possible to prepare a coupled BEC with
two different types of atoms.   In addition to coupled atomic condensates,  
there has been consideration of a hybrid BEC where one type of
bosons are  atoms and the other molecules \cite{cpl2}.  These initiated
theoretical
activities in BEC involving more than one types of bosons using the
coupled
GP equation \cite{cpl2,cpl1}.

In addition to just forming a coupled BEC with two quantum states of the
same atom, these studies also yielded crucial information about the
interaction among component atoms and measured the percentage of each
quantum states in the condensate \cite{excpl1,excpl2}. It has been found
that $^{87}$Rb atoms
have repulsive interaction in all three quantum states. Also, the strength
of repulsive interactions in $F=1, m=-1$ and $F=2,m=2$ states are
essentially identical. The interaction between an atom in the $F=1, m=-1$
state and another in the $F=2, m=2$ state is repulsive. As the change in
the $m$ value of a atomic quantum state does not correspond to a
substantial structural change, it is likely that such change would not
correspond to a large change in the atomic interaction. 

Here we 
study theoretically the collapse in a coupled BEC composed of two quantum
states 1 and 2 of a bosonic atom  using the coupled
time-dependent GP
equation. 
We motivate this study by considering two possible atomic states
of $^7$Li whenever possible. 
An experiment of collapse in a coupled BEC has not yet been realized but
could be
possible in the future. In the case of $^7$Li the interaction in state 1
is
taken to be attractive which is responsible for  collapse. 
Here there are three types of interactions denoted by the
scattering lengths $a_{ij}$, $i,j=1,2$, between states $i$ and $j$. A
negative (positive) scattering length denotes an attractive (repulsive)
interaction. We study the collapse with  different possibilities of
attraction and
repulsion between atoms in state 1 and 2. 
If one of the scattering lengths is negative, at least one
component of the condensate may experience collapse. If two of the
scattering lengths are negative one can have collapse in both components.
Specifically, one can also have collapse of both components if $a_{12}$ is
negative and $a_{ii}$, $i=1,2$ are positive.

The usual GP equation conserves the number of atoms.  The dynamics of the
collapse (growth and decay of number of atoms)  is best studied by
introducing an absorptive contact interaction in the GP equation which
allows for a growth in the particle number from an external source. One
has also to introduce an imaginary quartic three-body interaction term
responsible for recombination loss from the condensate \cite{2}.  If the
strengths of these two terms are properly chosen, the solution of the
time-dependent GP equation could produce a growth of the condensate with
time when the number of atoms is less than the critical number. Once it
increases past the critical number, the three-body interaction takes
control and the number of atoms suddenly drops below the critical level by
recombination loss signaling a collapse \cite{2}. Then the absorptive term
takes over and the number of atoms starts to increase again. This
continues indefinitely showing an infinite sequence of collapse.

\section{Coupled Gross-Pitaevskii Equation with absorption}

We consider the following spherically symmetric coupled GP equation with
two components at time $\tau$ for the condensate wave function
$\psi_i(r,\tau)$ \cite{cpl1}
\begin{eqnarray}\label{cc} \biggr[
-\frac{\hbar^2}{2m}\frac{1}{r}\frac{\partial^2 }{\partial
r^2}r &+& \frac{1}{2}c_im\omega^2 r^2 +\sum_{j=1}^2
g_{ij}N_j|\psi_j({
r},\tau)|^2 \nonumber 
\\
&-& \mbox{i}\hbar\frac{\partial}{\partial \tau}\biggr]
\psi_i(r,\tau)=0,
\end{eqnarray}
$i=1,2$, 
where  $m$ is the   atomic mass. Here
$g_{ij}=4\pi\hbar^2a_{ij}/m$ is the coupling constant for atomic
interaction, $N_j$ the number of condensed atoms in state 
$j$, and $\omega$ the frequency of the harmonic oscillator trap. The
parameter $c_i$ has been introduced to modify the frequency
of the trap for the atoms in each quantum state.  

As in Refs. \cite{4} it is convenient to use dimensionless variables
defined by $x = \sqrt 2 r/a_{\mbox{ho}}$ , and $t=\tau \omega, $
where
$a_{\mbox{ho}}\equiv \sqrt {\hbar/(m\omega)}$, and $
\phi_i(x,t) = x\psi_i(r,\tau ) (\sqrt 2\pi a_{\mbox{ho}}^3)^{1/2}$ . In
terms of these
 variables Eq. (\ref{cc})  becomes \cite{4}
\begin{eqnarray}\label{e}
\biggr[ -\frac{\partial^2 }{\partial
x^2} &+& \frac{ c_ix^2}{4} +\sum_{j=1}^2 n_{ij}
\frac{|\phi_j({x},t)|^2}{x^2}
-\mbox{i}\xi_i\frac{|\phi_i({x},t)|^4}{x^4}\nonumber \\
&+&\mbox{i}\gamma_i
-  \mbox{i}\frac{\partial
}{\partial t} \biggr]\phi_i({ x},t)=0, \end{eqnarray}
where
$n_{ij}\equiv 2\sqrt 2 N_j a_{ij}/a_{\mbox{ho}}$ 
 could be negative (positive) when the
corresponding interaction 
is attractive (repulsive). 
In Eq. (\ref{e}) we  have introduced
a diagonal absorptive i$\gamma_i$ and a quartic three-body term $
-\mbox{i}\xi_i{|\phi_i({x},t)|^4}/{x^4}$
appropriate to study
collapse \cite{2}.
For $\gamma_i=\xi_i=0, i=1,2$, the normalization condition of the wave
function is \begin{equation}\label{5}  \int_0 ^\infty |\phi_i(x,t)|
^2 dx = 1.  \end{equation} 
The 
root-mean-square (rms)
radius of the component $i$  $x^{(i)}_{\mbox{rms}}(t)$ at time $t$ is
defined by \begin{equation}\label{7}
x^{(i)}_{\mbox{rms}}(t)=
\left[\frac {\int_0
^\infty x^2 |\phi_i(x,t)| ^2 dx} {\int_0
^\infty  |\phi_i(x,t)| ^2 dx
}\right]^{1/2}.  \end{equation}

\section{Numerical results}

To solve Eq. (\ref{e})  we  discretize it in
both  space (using step 0.0001) and time (using step 0.05) employing a
Crank-Nicholson-type rule and reduce it to a
set of algebraic equations which
is then  solved by iteration using the known  boundary
conditions,
e.g., $|\phi_i(0,t)|=0,$ and $\lim_{x
\to \infty} |\phi_i(x,t)|\sim \exp(-x^2/4). $ 
The iteration is
started with the known normalized 
(harmonic oscillator) solution of Eq. (\ref{e}) obtained with $n_{ij}=0$
at $t=0$.  
The nonlinear constants $n_{ij}$ in this equation are  increased by
equal amounts 
over  500 to 1000 time iterations starting from zero until the desired
final
values are
reached.  
This iterative method  is  similar to one in the uncoupled case
\cite{2,4}. A
detailed account of the numerical procedure  for  the coupled case will
appear elsewhere.

\subsection{Stationary Problem}

First we  consider the stationary solution of Eq. (\ref{e}) with
$\gamma_i=\xi_i=0$, which illustrate the collapse. 
As the three scattering lengths $a_{ij}$ and two numbers $N_i$ are all
independent, the four parameters $n_{ij}$ are also so with one
restriction: the signs of $n_{12}$ and $n_{21}$ are identical.

Now we   study  the simplest case of
collapse by taking only the interaction between the  atoms  in state 
1 to be attractive corresponding to a negative $a_{11}$. All other
scattering lengths $-$
$a_{22}$ and $a_{12}$ (= $a_{21}$) $-$ are taken to be positive.  
Quite expectedly, here 
the first component of the condensate could experience collapse. 
Although the present formulation is generally valid, one has to choose
numerical values of the parameters before an actual calculation.  

The collapse of the first component  is illustrated in Fig. 1 (a) for $
n_{11}=-3.814, n_{22}=4$ $n_{12}=n_{21}=1, $
$c_1=0.25$, $c_2=4$. These parameters are in dimensionless units and one
can associate them with an actual physical problem of experimental
interest. For this we consider the state 1 to be the states of $^7$Li with
attractive interaction as in the actual collapse experiment with
$|a_{11}|/a_{\mbox{ho}}\simeq 0.0005$ \cite{1a}. As $n_{11}=2\sqrt 2 N_1
|a_{11}|/a_{\mbox{ho}}$ this corresponds to a boson number $N_1 \simeq 
2700. $ This number is larger than the maximum number  atoms
permitted in the  BEC of  a single component $^7$Li which is about 1400
\cite{1a}. The presence of the
second component with repulsive interaction allows for a formation of a 
stable BEC with more
$^7$Li atoms in quantum state 1 than allowed in the single-component BEC.  
Similar conclusion was reached by Esry \cite{esry} in a study of a coupled
BEC in a different context.
We find from Fig. 1 (a) that  $\phi_1$ is very
much centrally peaked compared to $\phi_2$. This corresponds to a small
rms radius and large central density for $\phi_1$ denoting an
approximation to  
collapse. If the number  $N_{1}$ is slightly increased beyond 2700 the
first
component of the condensate wave function becomes singular at the origin
and no
stable stationary solution to Eq. (\ref{e}) could be obtained.

Next we  discuss the collapse by taking only the interaction among 
atoms in two different states to be attractive corresponding to a negative
$a_{12}$ ($=
a_{21}$). The atomic interaction in both quantum states 1 and 2 is taken
to be repulsive corresponding to a  positive $a_{11}$ and $a_{22}$.
Although it is a problem of theoretical interest for the study of
collapse, it  has no experimental analogue in terms of $^7$Li.
We
illustrate
the approximation to collapse in this case in Fig. 1 (b) for parameters
$n_{11}=1, n_{22}=1.5, $ $n_{12}=-5.95,n_{21}=-2, $ $c_1=1$, $c_2=0.25$.
Both wave-function components are  peaked near $x=0$
and have small rms radii. The system would collapse with a small increase 
of
$|n_{12}|$ and/or $|n_{21}|$. Here the interactions among   atoms in
states
1 and 2 
are both repulsive. The collapse is a
consequence of the attraction between an atom in state 1 and one in state
2. This leads to a 
dominance of nonlinear off-diagonal coupling terms in the coupled GP
equation.

Finally, in Fig. 1 (c) we  illustrate the approximation to collapse of
both  components when all  scattering lengths are 
negative. This corresponds to taking all possible interactions attractive. 
The parameters in this case are $n_{11}=n_{22}=-1, $
$n_{12}=n_{21}=-0.552, $ $c_1=4$, $c_2=0.25$. This has an experimental
analogue in terms of two states of $^7$Li. We assume the atomic
interaction in both
states  to be equally attractive corresponding to a negative scattering
length: $a_{11}=a_{22}$. For
$|a_{11}|/a_{\mbox{ho}}\simeq 0.0005$ as in the actual experiment
\cite{1a}, one has 
$N_1 = N_2 \simeq 700.$  The total number of particles in this case is
roughly 1400, which is equal to the critical number observed in the actual
experiment of collapse in $^7$Li.
Both wave-function
components could become singular in this case as all possible interactions
are attractive. 

\subsection{Time-dependent Problem}

Although the collapse of the coupled condensates could be inferred from
the shape of the stationary wave functions of Fig. 1 (sharply peaked
centrally with small rms radii), we  also study the dynamics of
collapse from a time evolution of the full GP
equation (\ref{e}) in the presence of an absorption and three-body
recombination, e.g., for $\gamma_i \ne 0$ and $\xi_i \ne 0$ as in the
uncoupled case \cite{2}. For this purpose we  consider the solution of Eq.
(\ref{e}) normalized according to Eq. (\ref{5}) at $t=0$ obtained with
$\gamma_i=\xi_i=0$ and allow this solution to evolve in time with
$\gamma_i
\ne 0$ and $\xi_i \ne 0$ by iterating the GP equation (\ref{e}). The
fractional change in the number of atoms due
to the combined effect of absorption and three-body recombination is given
by 
\begin{equation} \frac{N_i(t)}{N_i(0)} = \frac{\int_0 ^\infty
|\phi_i(x,t)|^2 dx}{\int_0 ^\infty |\phi_i(x,0)| ^2 dx} \quad ,
\end{equation}
and the rms radii by Eq. (\ref{7}). The continued growth and decay of the
number of particles in the condensate would signal  the possible collapse
in a particular case. The oscillation of the rms radius would  demonstrate
the consequent radial vibration of the condensate.

Now we  study the time evolution of the number of atoms of the two
components and the corresponding rms radii. The general
nature of time evolution is independent of the actual values of $\gamma_i$
and $\xi_i$ employed provided that a very small value for $\xi_i (\sim
0.001)$ and a relatively larger one for $\gamma_i (\sim 0.01$ to 0.1)  are
chosen \cite{2}. The following parameters were chosen in case of models
(a), (b),
and (c) of Fig. 1: (a) and (b)  $\gamma_1 =\gamma_2= 0.03,
\xi_1=\xi_2=0.001$, (c) $\gamma_1 =0.15, \gamma_2= 0.03, \xi_1=0.002,
\xi_2=0.003$.  The fractional change in the
number of atoms for the two components are shown in Figs. 2 (a), (b), and
(c). The results for $0<t<100$ in Fig. 2 are calculated with 2000
iterations of the GP equation (\ref{e}) using a time step 0.05.

The quadratic nonlinear terms in model (a) are all repulsive in channel 2,
the corresponding wave function ($\phi_2$) of Fig. 1 (a) does not show any
sign of approximation to collapse as in channel 1 where the diagonal
nonlinear term is attractive. The results reported in Fig. 2 (a) are
consistent with this. The number of particles $N_1$ of the first component
undergoes successive growth and decay, whereas that of the second
component keeps on growing indefinitely typical to a repulsive
interaction.

For model (b) the effective nonlinear terms in channels 1 and 2 are both
repulsive and it should be possible to have  collapse in both channels
by decreasing the dominating off-diagonal quadratic nonlinear terms
$n_{12}$ and $n_{21}$ corresponding to an increase in attraction between
an atom in
state 1 and one in state 2. However, for the actual parameters of this
model
only the component 1 exhibits collapse. This is consistent with the more
singular nature of $\phi_1$ reported in Fig. 1 (b), compared to $\phi_2$.
Consequently, in Fig. 2(b) only component 1 experiences collapse; the
number of particles $N_2$ keeps on growing with time.

In model (c) all the quadratic nonlinear terms are
attractive. Consequently, in Fig. 2 (c)  we  find a series of collapse
in both channels.  The collapse is most favored in model (c)  with
attractive diagonal and nondiagonal nonlinear terms. This corresponds
to attraction between two atoms in state 1, between two atoms in state
2, and between an atom in state 1 and another in state 2. 
The next favored case
is of model (a) where the diagonal nonlinear term is negative in channel
1. Here only the atomic interaction in state 1 is attractive, all other
atomic interactions are repulsive. 
The least favored case is of model (b) where only the off-diagonal
nonlinear terms are negative.  This corresponds
to repulsion  between two atoms in state 1, and between two atoms in state
2, and attraction between an atom in state 1 and another in state 2.
In the last case, collapse takes place due
to the dominance of the attractive nondiagonal nonlinear term over the
repulsive diagonal one
in channel
1.  This is explicit in Fig. 2 where the frequency of collapse decreases
from model (c) to (a) and then to (b).

Finally, in Figs. 3 (a), (b), and (c) the rms radii for the two
components are shown for  models of Figs. 2 (a), (b), and (c),
respectively. In case of models (a) and (b) we  find from Figs. 2 (a) and
(b) that the number $N_2$ grows with time. This is
reflected in the growth of the corresponding  rms radii in Figs. 3 (a)
and (b). 
In case of model (c) there is collapse in both channels and both the rms
radii oscillate with time. This radial vibration of the collapsing
condensate(s) also takes place in the uncoupled case \cite{2}. However,
from Figs.
3 (a) and (b) we  find that due to a collapse in one of the channels, both
rms radii could execute oscillations. In one of the channels it is a
direct
consequence of collapse, in the other it is due to a coupling to the
channel experiencing collapse.    

\section{Conclusion}

To conclude, we studied the collapse in a trapped BEC of atoms in states 1
and 2 using the GP equation when some of the atomic interactions are
attractive. We motivate parts of this study with two atomic states of
$^7$Li. The component $i$ of the condensate could experience collapse when
the interaction among atoms in state $i$ is attractive.  Both components
could experience collapse when at least the interaction between an atom in
state $1$ and one in state $2$ is attractive.  The collapse is predicted
from a stationary solution of the GP equation. The time evolution of
collapse is studied via the time-dependent GP equation with absorption and
three-body recombination.  The number of particles of the component(s) of
BEC experiencing collapse alternately grows and decays with time.  With
the possibility of observation of coupled BEC, the results of this study
could be verified experimentally in the future.

The work is supported in part by the Conselho Nacional de Desenvolvimento
Cient\'\i fico e Tecnol\'ogico  and Funda\c c\~ao de Amparo \`a Pesquisa
do Estado de S\~ao Paulo of Brazil.

\newpage

{\bf Figure Caption:}

1. Wave function components $\phi_1(x)$ (full line) and $\phi_2(x)$
(dashed line) vs. $x$ for two coupled GP equations with (a)  $
n_{11}=-3.814, n_{22}=4$, $n_{12}=n_{21}=1, $ $c_1=0.25$, $c_2=4$;  (b)
$n_{11}=1,
n_{22}=1.5, $ $n_{12}=-5.95,n_{21}=-2, $ $c_1=1$, $c_2=0.25$; and (c)
$n_{11}=n_{22}=-1, $ $n_{12}=n_{21}=-0.552, $ $c_1=4$, $c_2=0.25$.

2. The fractional change in the number of atoms $N_i(t)/N_i(0)$ vs. $t$
for component 1 (full line) and 2 (dashed line)
 for models (a) and (b) with $\gamma_1=\gamma_2=0.03 $ and 
$\xi_1=\xi_2=0.001$, and for (c) with $\gamma_1= 0.15,
\gamma_2=0.03$,
$\xi_1= 0.002$, and $\xi_2=0.003$. The parameters are  as
in Fig. 1.

3. The time dependence of rms radii $x_{\mbox{rms}}^{(i)}(t)$ of models
(a), (b), and (c) for 
 component 1 (full line) and 2 (dashed line). The
parameters are as in Figs. 1 and 2.

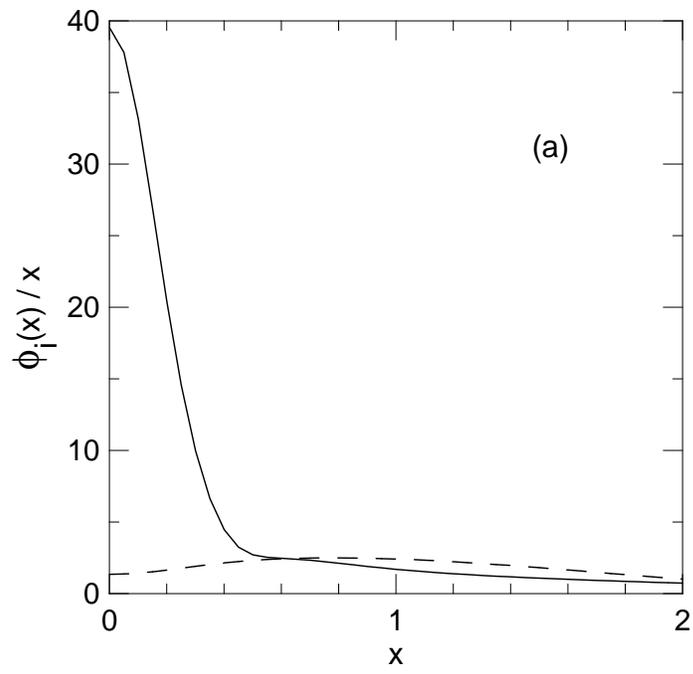
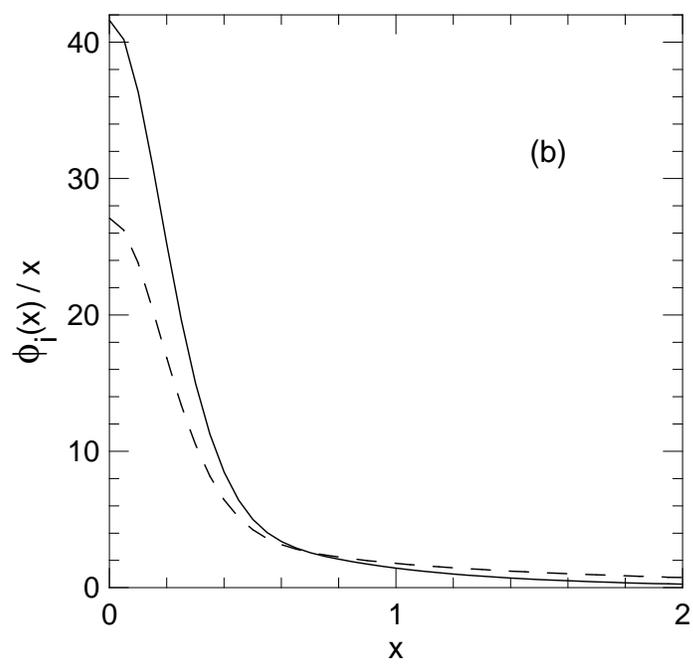
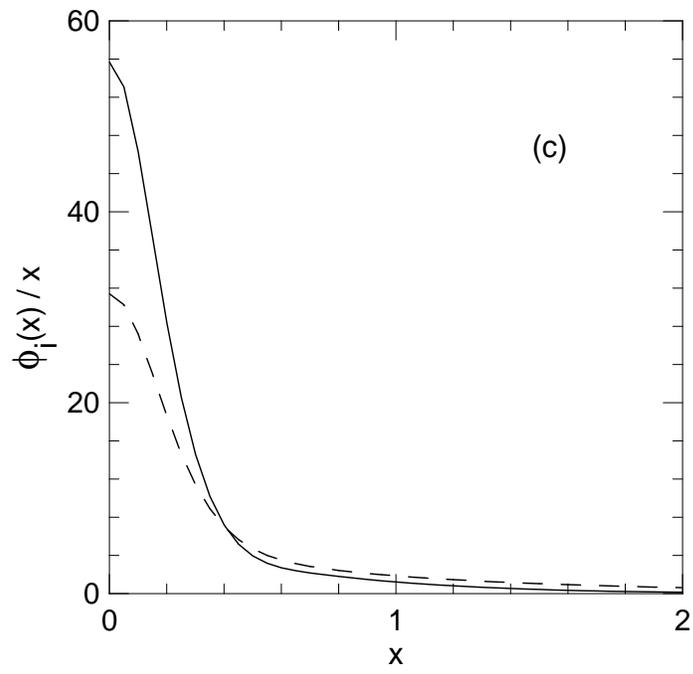
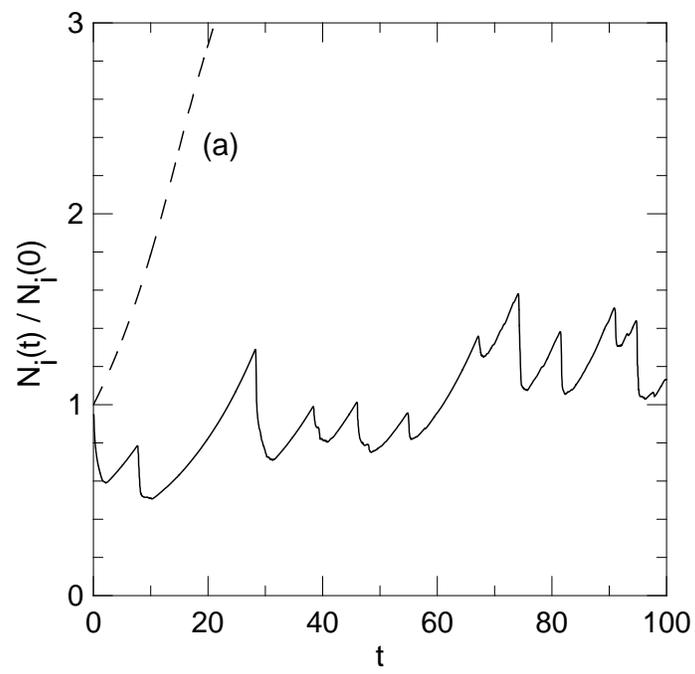
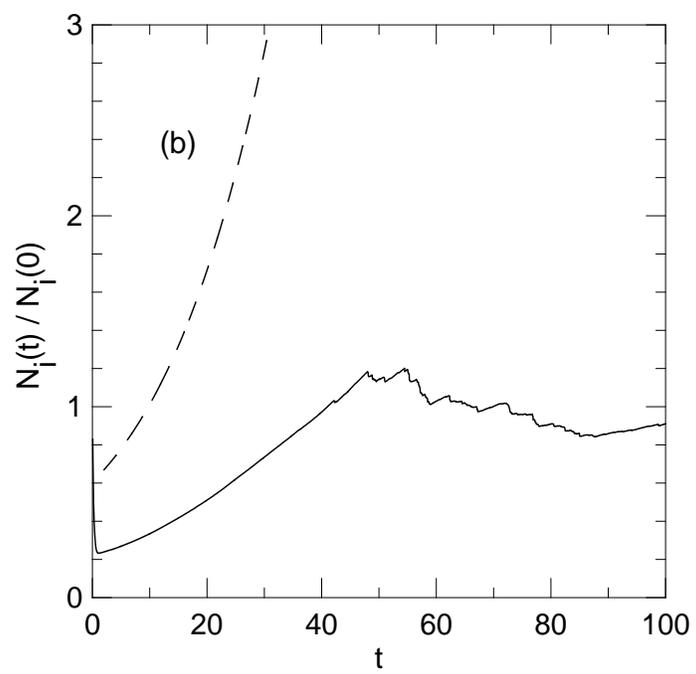
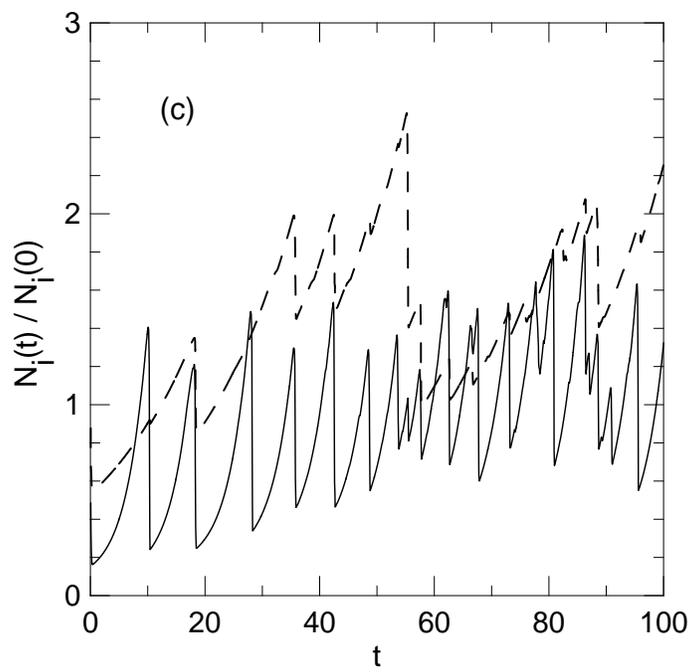
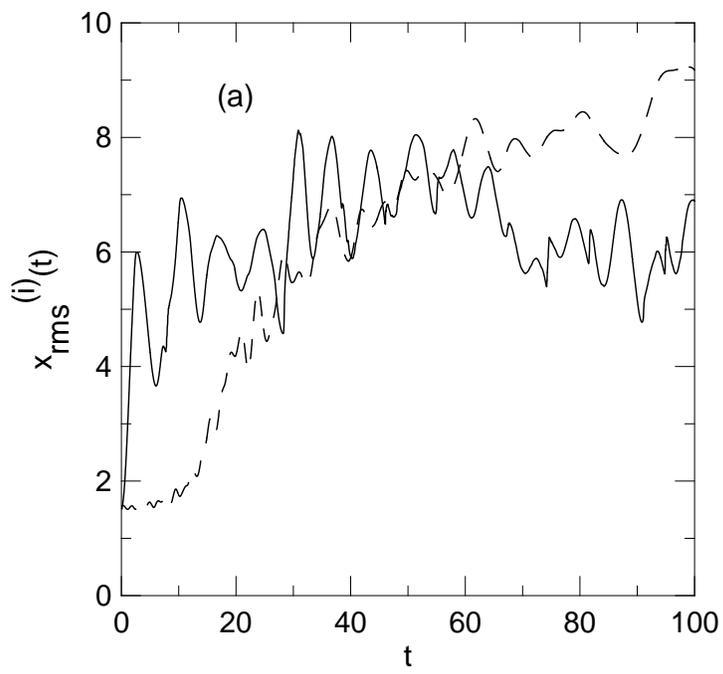
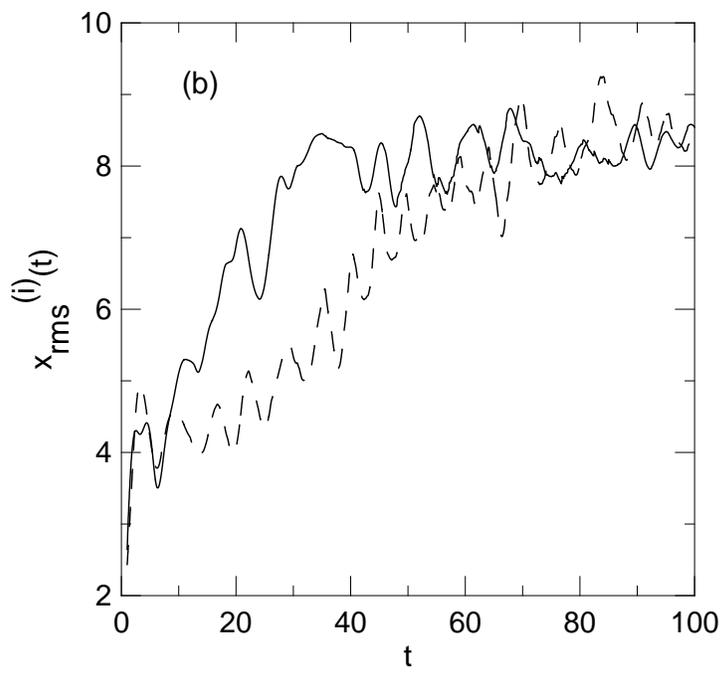
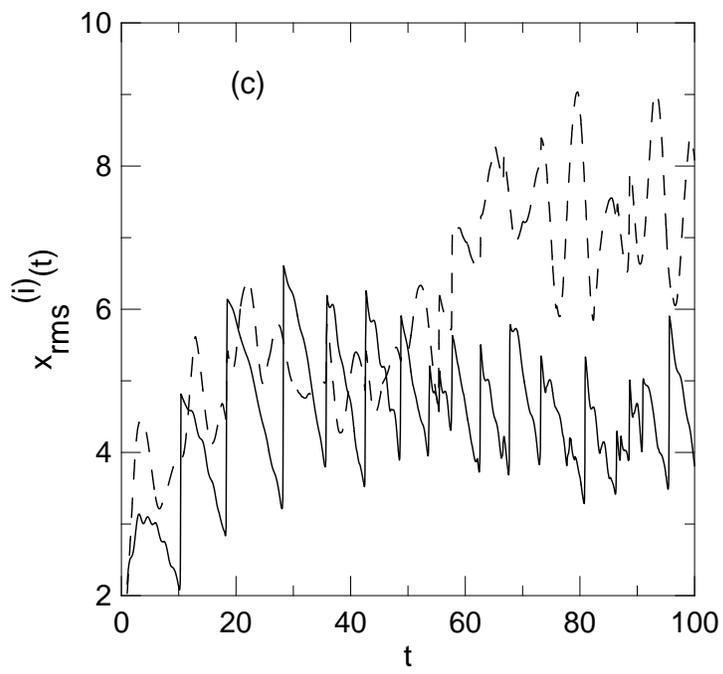
\end{document}